\documentclass[
 reprint,
showpacs,
 amsmath,amssymb,
 aps,
pra,
]{revtex4-1}

\usepackage{graphicx}
\usepackage{dcolumn}
\usepackage{bm}
\usepackage{hyperref}

\def\v#1{{\bf#1}}
\def\be{\begin{equation}}
\def\ee{\end{equation}}
\def\bea{\begin{eqnarray}}
\def\eea{\end{eqnarray}}
\def\ahalf{{\textstyle{1\over2}}}

\def\hcal{\mbox{$\cal H\,$}}
\def\ncal{\mbox{$\cal N\,$}}
\def\<{\langle}
\def\>{\rangle}

\begin{document}


\title{Transparent lattices and their solitary waves}

\author{E. Sadurn\'i}
 \email{sadurni@ifuap.buap.mx}
\affiliation{Instituto de F\'isica, Benem\'erita Universidad Aut\'onoma de Puebla,
Apartado Postal J-48, 72570 Puebla, M\'exico}

\date{\today}

\begin{abstract}
We provide a familiy of transparent tight-binding models with non-trivial potentials and site-dependent hopping parameters. Their feasibility is discussed in electromagnetic resonators, dielectric slabs and quantum-mechanical traps. In the second part of the paper, the arrays are obtained through a generalization of supersymmetric quantum mechanics in discrete variables. The formalism includes a finite-difference Darboux transformation applied to the scattering matrix of a periodic array. A procedure for constructing a hierarchy of discrete hamiltonians is indicated and a particular biparametric family is given. The corresponding potentials and hopping functions are identified as solitary waves, pointing to a discrete spinorial generalization of the Korteweg-deVries family. 
\end{abstract}



\maketitle


\section{ Introduction \label{sec:1}}

There is an increasing interest in the study of tight-binding models and their application to several areas, such as optical waveguides \cite{russell2003, keil2010, foresi1997}, ultracold atoms in optical lattices \cite{oberthaler2006, bloch2005, oberthaler1996} and artificial realizations of condensed matter, e.g. graphene \cite{esslinger2013, kuhl2010, barkhofen2013, bittner2010, bellec2013}. Noteworthy is the area of quantum metamaterials \cite{felbacq2012, luukkonen2009}, where transmission properties can be tailored at the level of quantum degrees of freedom. The subject has reached the domain of supersymmetric models through a recent realization of a Dirac oscillator \cite{franco2013, sadurni2010, longhi2010}, which seems to be the first experimental construction of an $N=2$ supersymmetry. In these applications, technical developments towards the engineering of local potentials and couplings between sites have been reached with several purposes \cite{esslinger2008, christodoulides2008}, even with microwaves \cite{sadurni2013}. A particular example of interest is the perfect transmission of signals through discrete arrays and its associated scattering problem in the presence of local lattice modifications. It is important to mention that transparency has been extensively studied in continuous variables \cite{cooper1995, samsonov2000}, but never in a tight-binding array \cite{berry2008}. In this paper we provide an elegant solution of the problem by extending the well-developed apparatus of supersymmetric quantum mechanics \cite{kac1973, catterall2002} (SUSYQM) to discrete variables. 

We shall proceed in the following order: In section \ref{sec:2} we define our problem and state the main result of our work by giving a family of transparent potentials in tight-binding arrays. In subsection \ref{sec:experiment} we study the possibility of implementing such models: matter waves in optical traps \ref{sec:qmwaves}, electromagnetic waves in nanoscopic and mesoscopic arrays \ref{sec:emwaves} and dielectric slabs \ref{sec:slabs} are considered. Section \ref{sec:3} provides the necessary definitions and generalizations for discrete SUSYQM. This section can be read separately and is divided in five parts: \ref{sec:3.0} The factorization method, \ref{sec:3.1} isospectrality and transparency (where we address the problem of discrete-variable Darboux transformations), \ref{sec:3.2} the continuous limit of the theory, \ref{sec:4} a numerical check of the required properties and, in full analogy with traditional SUSYQM, subsection \ref{sec:5} studies a set of discrete solitons given by potentials and hopping parameters.  Section \ref{sec:6} gives a brief conclusion. 

\section{A family of models \label{sec:2}}

\begin{figure}[t]
\includegraphics[width=8.6cm]{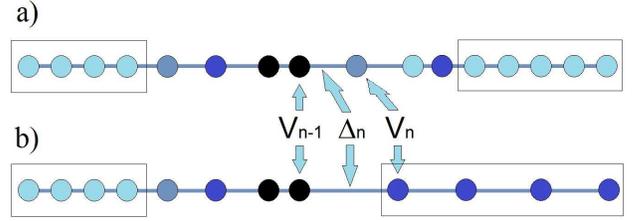} 
\caption{\label{fig:0} Two asymptotically periodic arrays. Bonds represent couplings $\Delta_n$ and site colors represent potential values $V_n$. Example a) has two identical asymptotic regions indicated by rectangular boxes. In example b), the asymptotically periodic regions do not coincide.}
\end{figure}

\begin{figure*}[t]
\includegraphics[width=10.6cm]{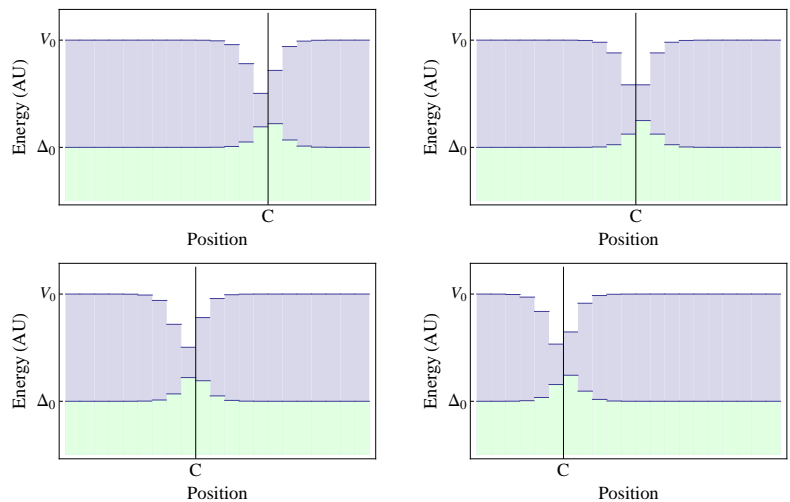}
\caption{\label{fig:1} Variation of our discrete potential (blue-filled curve) and hopping function (green-filled curve) with respecto to $\gamma$. The minimum of the potential and the maximum of the hopping parameter lie in the same region and suffer a linear displacement to the right when $\gamma$ is increased exponentially (the center is denoted by C).}
\end{figure*}

\begin{figure}[t]
\includegraphics[width=7cm]{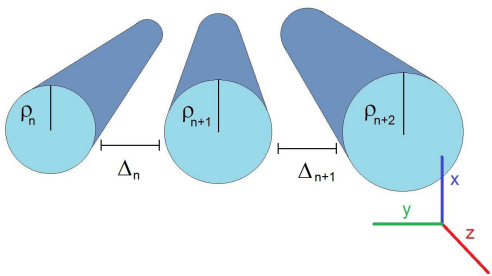}
\caption{\label{fig:1.1} Dielectric rods of variable radii. The couplings $\Delta_n$ can be tuned by varying the distance between the cylinders. The optical axis $z$ is parallel to their longitudinal coordinate.}
\end{figure}

\begin{figure}[t]
\includegraphics[width=7cm]{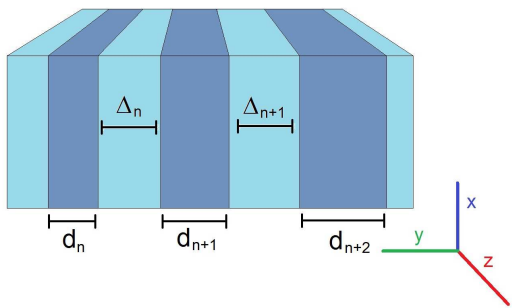}
\caption{\label{fig:1.2} Dielectric slabs of variable thickness and alternating refraction index. As before, the distance between dark-blue slabs can be used to control $\Delta_n$.}
\end{figure}

Several areas benefit from second quantized tight-binding models with and without particle interactions. In one dimension, and with units such that $\hbar=1$, one typically has

\bea
H_{\mbox{\small quantum}} = \sum_{\< n, m\>} \left[\Delta_{nm} C^{\dagger}_n C_m +  \Delta_{nm}^*  C^{\dagger}_m C_n  \right]
\label{a1}
\eea
where $C_n, C_m^{\dagger}$ are local field operators of bosonic or fermionic nature. The off-diagonal $\Delta_{nm}$'s are hopping parameters and the diagonal $\Delta_{nn}$ represent on-site potentials. The propagation of signals in such models can be viewed as a scattering problem in which one or many particle states with well defined momenta can be ideally prapared at infinity. In the absence of terms of the form $C_n^{\dagger} C_m C_i^{\dagger} C_j$ (dilute matter waves \cite{zoller1998} or photonic crystal waveguides \cite{chien2007}), the lack of particle interactions allow a single-particle treatment of the scattering problem. Therefore, we restrict ourselves to a hamiltonian in first quantization. With the help of Wannier functions $\< x | n \>$ localized around site $n$, we have

\bea
H = \sum_{n} \Delta_n | n \>\<n-1| + \Delta_n^* | n-1 \>\<n| + V_n | n \>\<n|.
\label{a2}
\eea 
For simplicity, only nearest neighbors are assumed. One can expand a stationary state $|\psi \>$ as

\bea
|\psi \> = \sum_{n} \psi_n | n \>
\label{a3}
\eea
and from this expansion, the Schr\"odinger equation associated to (\ref{a2}) acquires its typical recurrence form

\bea
\Delta_n \psi_{n-1} + \Delta_{n+1} \psi_{n+1} + V_n \psi_n = E \psi_n.
\label{a4}
\eea
Now we define $H$ to be {\it asymptotically periodic\ }if at the far ends of the array (left and right) we have asymptotically constant couplings and potentials, i.e. $\Delta_{n} \sim \Delta_0$ and $V_n \sim V_0$  if $n >> 1$ . A diagram is shown in figure \ref{fig:0}. The main finding of this work is the existence of a family of potentials and hopping paramaters (or functions, since they depend on site $n$) for which the reflection coefficient of a Bloch wave vanishes for all energies. The most general expression for the biparametric family can be written in terms of continued fractions, as we shall see in section \ref{sec:5}. To this end, we employ the notation

\bea
\left[a_n, b_n; a_{n-1}, b_{n-1}; ... \right] = a_n - \frac{b_n}{a_{n-1} - \frac{b_{n-1}}{\cdots}}.
\label{a6}
\eea
An important case is given by a monoparametric subfamily $\tilde V_n, \tilde \Delta_n$, which can be written in terms of hyperbolic functions

\bea
\tilde V_n &=& V_0  \nonumber \\ &+& \Delta_0\left(\frac{\cosh\left[ n \lambda + \beta \right]}{\cosh\left[ (n-1) \lambda + \beta \right]} - \frac{\cosh\left[ (n+1) \lambda + \beta \right]}{\cosh\left[ n \lambda + \beta \right]}\right), \nonumber \\ 
\label{a7}
\eea
\bea
\tilde \Delta_n = \Delta_0 \frac{\sqrt{ \cosh \left[ (n-2) \lambda + \beta \right] \cosh \left[ n \lambda + \beta \right] } }{ \cosh \left[ (n-1) \lambda + \beta \right] },\nonumber \\ 
\label{a8}
\eea
where $\beta$ is a free parameter and the convenient definition 
\bea
\lambda = \ahalf \ln \left(\frac{V_0-\sqrt{V_0^2 - 4 \Delta_0^2}}{V_0+\sqrt{V_0^2 - 4 \Delta_0^2}}\right) 
\label{a9}
\eea
has been used. Evidently, $|V_0| \geq 2|\Delta_0|$ is a necessary condition and it can always be met by recognizing that $V_0, \Delta_0$ are independent parameters. Some examples are depicted in figure \ref{fig:1} as a function of $\gamma=\mbox{e}^{2\beta}$.

\subsection{Experimental feasibility \label{sec:experiment}}

Tight binding models can be implemented in a variety of settings, ranging from matter waves to electromagnetic waves. The implementation of our model relies strongly on the possibility of reaching physical values for our on-site energies $V_n$ and hopping functions $\Delta_n$. In this respect, it is important to ensure that all the values provided by (\ref{a7}, \ref{a8}) are bounded. It is a simple task to find upper and lower bounds for such functions:

\bea
|\Delta_0| < |\tilde \Delta_n| < |\Delta_0 \cosh (\lambda)|, 
\label{b1}
\eea
and
\bea
V_0 + \Delta_0 \left( 1 - \frac{\cosh(3\lambda/2)}{\cosh(\lambda)} \right) < \tilde V_n < V_0 + \Delta_0. 
\label{b2}
\eea
These bounds depend entirely on $V_0, \Delta_0$, and there is no value of $\lambda$ in (\ref{a9}) that leads to singular limits for non-trivial configurations.

With these bounds, we are in the position to discuss realistic parameters in available experiments.

\subsubsection{Quantum-mechanical waves \label{sec:qmwaves}} With the aim of understanding wave propagation in solids, artifical crystals have been produced by means of optical traps. Bose-Einstein Condensates (BECs) are produced and held by potentials that, in some applications, can be tuned in frequency and shape. Hubbard models with tailored hopping parameters \cite{struck2012} have been achieved with different purposes, such as the emulation of non-abelian gauge fields in 2d (phase variation of $\Delta$) and the study of localization phenomena, with its corresponding insulator transition as a function of disorder \cite{inguscio2007}. It is important to recall that reasonable on-site energies have values around $|V_n-V_0|<h \times 3.2$kHz, providing thus an interval to play with. The hopping parameters are in turn determined by such potential heights and by the lattice spacing ($830$ nm $<\lambda<1076$ nm for laser traps). Moreover, the minimization of self-interactions (or supression of non-linear terms in the effective Gross-Pitaevskii wave equation) has been achieved in BECs with atoms such as K, Cs and Li, making linear tight-binding models more realistic \cite{roati2007}. We must recognize, however, that full trap tunability depends on the use of lasers at many different frequencies. Well widths and depths can be controlled in this way, but recent applications have only used a few frequencies; for example Ref. \cite{inguscio2007} reports two colors in order to produce non-periodic configurations.       

In a different domain of quantum mechanics, we find the so-called quantum metamaterials \cite{felbacq2012}, where it is possible to tune the transmission properties of arrays made of doped nanorods. By illuminating our proposed configuration of resonators, one may be able to switch its properties from full reflectivity (due to a bandgap) to transparency (as in our case). 

\subsubsection{Electromagnetic waves \label{sec:emwaves}} We propose the use of dielectric media disposed in convenient configurations (figs. \ref{fig:1.1} and \ref{fig:1.2}), such as those employed in the fabrication of photonic crystals (nanoscopic scales) and microwave resonators (mesoscopic scales). The first possibility is within range, if we recall that carved structures in controlled patterns have been consistently produced for a few decades \cite{yablanovitch1987, little1995}. See \cite{bienstman2003} and, in particular, a so-called {\it stage I coupler\ }in the inset of fig. 5 of the same paper. The main idea is to vary the size, the interdistance and refraction indices of such structures. Realistic parameters can be quoted; for example, in \cite{little1995} we find refraction indices in the range $3.20<n<3.25$ and slab thicknesses $d\sim 1\mu$m. Realizations with nanorods of AlO and GaAs mixtures work with $1.61<n<3.37$, as reported in \cite{bienstman2003}, and with variations of their radii (around $\rho \sim 125$nm) one may further induce shifts in their resonant frequencies. In this way, one may achieve a variable on-site potential in effective tight-binding models as a function of resonator size. The hopping parameters can be tuned by varying the distance between structures such as slabs, rods or coupled waveguides in general. Finally, the values cited above determine the elements of transfer matrices for the propagation of TE and TM modes, e.g. Eq. (1) in Ref. \cite{wang2008}. 
 
For photonic structures carved in Si substrates \cite{wang2008}, the reported permittivities are $\epsilon_a = 2.22$ and $\epsilon_b =1$ (air layers). Frequency gaps can be achieved at central wavelengths $\lambda_{\scriptsize \mbox{center}} \sim 1011$ nm for normal incidence of light, but such values can be tuned down to $350$ nm by varying the angle of incidence, covering the complete optical range. When the structures are built in a non-periodic pattern of a similar size, the corresponding frequency levels tend to be discretely distributed, but they lie in the same (optical) region. The thickness of each barrier or slab can be in the range $95$ nm $<d<459$ nm.

Other realizations with variable couplings using photons in waveguides have been reported \cite{keil2013}. The couplings $\Delta_n$ are tuned again by the interdistance mechanism. Similar settings were produced in \cite{sadurni2013}. An exponential law for $\Delta_n$ as a function of interdistance $d$ was shown in both cases. This dependence can be further refined by modified Bessel functions arising from cylindrical geometries. 

In the case of ceramic microwave resonators  (Temex series from E2000 to E7000), different types of spectra have been produced by the method of variable couplings \cite{kuhl2010, franco2013, sadurni2013}. The parameters were tuned by analyzing the level separation of a dimer as a function of the distance between resonators. It is important to mention that the emulation of (\ref{a7}) also requires a mechanism that adjusts the on-site potential. To this end, it is convenient to employ resonators made of different materials, with selected resonant frequencies ranging from 800MHz to 50GHz and various permittivities. Isolated sharp peaks of $\Gamma \sim$2MHz for each resonator provide reasonable spectra in arbitrary arrays. It should be noted that other approaches to vary on-site energies are possible, as indicated in \cite{barkhofen2013}. The spectral gap of Boron Nitride was emulated by breaking a dimer symmetry using antenna couplings, rather than varying the permittivity of the constituents. 

\subsubsection{Couplings and on-site potentials for dielectric layers \label{sec:slabs}}

Let us analyze a problem of parallel dielectric slabs, with the aim of extracting the thickness and interdistance dependence of couplings and on-site potentials. We start with a single slab of permittivity $\epsilon_1$ immersed in a medium of permittivity $\epsilon_2$ (cladding). We have $\epsilon_1>\epsilon_2$. If the slab is parallel to the $x$ axis, as in fig. \ref{fig:1.2}, we will have a continuous TE mode $\v E_x$ travelling along $\v k$, which lies in the $y$-$z$ plane. We would like to analyze the behavior of $\v E_x$ along the coordinate $y$, i.e. across media interfaces. Our starting point is the 3d Helmholtz equation ($c=1$)

\bea
\left( \nabla^2 + \epsilon_1 \omega^2 \right) \v E_x = 0, \quad \v r \in \Omega, \nonumber \\
\left( \nabla^2 + \epsilon_2 \omega^2 \right) \v E_x = 0, \quad \v r \notin \Omega.
\label{b3}
\eea 
This equation can be transformed to a Schr\"odinger-like 1d equation with a potential well: assuming no confinement along $x$, we reduce (\ref{b3}) to 

\bea
\left( -\frac{\partial^2}{\partial y^2} + U_0 \Theta(L/2 - |y|) \right) \phi(y) = E \phi(y)
\label{b4}
\eea
with $E\equiv\epsilon_2 \omega^2 - k_z^2 <0, U_0 \equiv (\epsilon_2 - \epsilon_1 )\omega^2 <0$ and $\v E_x(x,y,z)=\v x \phi(y) e^{ik_z z}$. The solutions $\phi(y)$ are confined along $y$ by the walls of the potential, i.e.

\bea
\phi(y) = \ncal \times \begin{cases}
\exp (-\frac{|y-L/2|}{\lambda}) & \text{($y > L/2$)} \\
\exp (-\frac{|y+L/2|}{\lambda}) & \text{($y < -L/2$)},
\end{cases}
\label{b5}
\eea
and for $|y|<L/2$, $\phi$ is a trigonometric function. The skin depth $\lambda$ is related to the effective energy by the simple relation $\lambda = 1/\sqrt{|E|}$. In order to find $E$ we must solve a well-known transcendental equation; with the variables $\xi_0 \equiv L^2 U_0 / 4, \xi \equiv \sqrt{E+U_0} L / 2$, one has $\sqrt{\xi_0^2-\xi^2} = \xi \tan \xi$, but in the regime of interest only one bound state is needed. Therefore we take $\xi \ll 1$ and solve the previous relation, which leads to the following estimate to lowest order in $U_0$:

\bea
E \approx -\frac{L^2  U_0^2}{4} = -\frac{ \omega^4 L^2  (\epsilon_1-\epsilon_2)^2}{4} 
\label{b6}
\eea
and
\bea
\lambda \approx  \frac{4}{\omega^2 L  (\epsilon_1-\epsilon_2)}. 
\label{b6.1}
\eea
We are now able to find the couplings of a nearest-neighbor hamiltonian. When a second slab of permittivity $\epsilon_1$, skin depth $\lambda'$ and width $L'$ is centered at $y=d +(L + L')/ 2$, a significant overlap between functions $\phi, \phi'$ appears and the eigenfrequencies of the system are modified. We obtain a non-diagonal element in the effective hamiltonian (\ref{b4}), given by

\bea
\Delta = \int_{-\infty}^{+\infty} dy (\phi H_{\scriptsize \mbox{eff}} \phi'^*) \approx  - \int_{L/2}^{L/2 + d} dy \phi \frac{ d^2 \phi'}{dy^2}.
\label{b7}
\eea
This integral contains exponential tails and it can be carried out easily. It is important to note that the normalization constants of $\phi, \phi'$, i.e. $\ncal, \ncal'$, must be redefined in the interval $[L/2,L/2+d]$. With this in mind, we obtain an exponential dependence

\bea
\Delta(d) &=& \Delta(0) \exp \left( -\frac{(\lambda + \lambda') d}{2 \lambda \lambda'} \right) \nonumber \\ &\times& \frac{\sqrt{\lambda \lambda'} \sinh \left( \frac{(\lambda - \lambda') d}{2 \lambda \lambda'} \right) }{(\lambda-\lambda')\sqrt{\sinh \left( \frac{d}{2 \lambda} \right)\sinh \left( \frac{d}{2 \lambda'} \right)} }.
\label{b8}
\eea
For practical purposes, we can approximate this expression by the exponential alone. On the other hand, the on-site potentials in a tight-bidning approximation are given by the eigenvalues of isolated potential wells $E \sim 1/\lambda^2, E' \sim 1/\lambda'^2$. Finally, we can see that a careful control of the independent parameters $L, L', d$ -- and possibly $\epsilon_1, \epsilon_2$ -- generates couplings and on-site potentials that can be used in arrays of many sites. Given a set of couplings and potentials $\{ \Delta_n, V_n \}$, we obtain
a set of widths and separations $\{ L_n, d_n \}$ through the relations
\bea
 L_n = \frac{2}{\omega^2(\epsilon_1-\epsilon_2)} \sqrt{|V_{\scriptsize \mbox{offset}}-V_n|},
\label{b9}
\eea
\bea
d_n = - \left( \frac{2}{\sqrt{|V_n|} + \sqrt{|V_{n+1}|}} \right) \log \left( \frac{\Delta_n}{\Delta(0)} \right). 
\label{b10}
\eea

\section{A path to discrete SUSYQM \label{sec:3}}

In the rest of this paper we establish the mathematical methods that lead to transparency in the context of discrete variables. This shall be done with the help of discrete SUSYQM. 

\subsection{The discrete factorization method \label{sec:3.0}}
 The central discretization of the Schr\"odinger equation leads naturally to a nearest-neighbor tight-binding hamiltonian, therefore we focus on such local quantum-mechanical models for our constructions. A generic expression for $H$ in terms of operators can be written as

\bea
H = \Delta(N) T + T^{\dagger} \Delta(N)^{\dagger} + V(N),
\label{e1}
\eea
where $N$ is the site number operator and $T$ is a translation in one unit. These operators satisfy $\left[ F(N), T \right]= T \{F(N+1)-F(N) \}$ for any function $F$. Their action on localized states $|n\>$ is given by $N|n\> = n |n\>$ and $T |n\> = |n+1\>$. The functions $V(N)$ and $\Delta(N)$ represent the on-site potential and the nearest- neighbor hopping function, respectively. Their eigenvalues are $V_n$ and $\Delta_n$, with $V_n$ real. In full analogy with traditional SUSYQM in continuous variables, we propose a factorization scheme of (\ref{e1}) as follows 

\bea
H = A^{\dagger} A, \quad A =  F(N)T + G(N).
\label{e2}
\eea
The reconstruction of (\ref{e1}) will be possible if we impose the restrictions $V_n = |G_n|^2 + |F_{n+1}|^2$ and $\Delta_n = G_{n}^* F_n$, as can be verified by applying the product $A^{\dagger} A$ to some state $|n\>$. It is important to recognize that this scheme can be applied only if the potential is positive definite or, without loss of generality, if it is bounded below. This property enables us to consider the existence of a ground state and to further subtract it from the hamiltonian, which is a usual procedure \cite{cooper1995}. We should also point out that given a positive $V_n$ and a complex $\Delta_n$, we may determine $F_n$ and $G_n$ up to phase factors (which can be gauged away trivially in 1d). To this end one has to solve the recurrence $V_n = |F_{n+1}|^2 + |\Delta_n|^2 / |F_n|^2$ for $|F_n|^2$ by the method of continued fractions

\bea
|F_n|^2 = \left[ V_{n-1}, |\Delta_{n-1}|^2;V_{n-2}, |\Delta_{n-2}|^2;... \right]
\label{e3}
\eea
and then reconstruct $G$ with the relation

\bea
|G_n|^2 = \frac{|\Delta_n|^2}{\left[ V_{n-1}, |\Delta_{n-1}|^2;V_{n-2}, |\Delta_{n-2}|^2;... \right]}.
\label{e4}
\eea

Our proposal for the factorization of $H$ is in full correspondence with the continuous case, in view of the analogy $ A_{\mbox{\small discrete}} \leftrightarrow  A_{\mbox{\small continuous}}$, i.e.

\bea
F(N) T + G(N) \longleftrightarrow \frac{d}{dx} + W(x),
\label{e4}
\eea
where $W(x)$ is a superpotential satisfying the Riccati equation \cite{cooper1995}. A clear connection with continuous variables can be given by means of a lattice spacing $a$: The translation of wave functions $\<x-a|n\>=\<x|n+1\>$ motivates the substitutions $T=\exp (- a \cdot d/dx),  x= aN$, which in turn lead to the limit   

\bea
F \left(\frac{x}{a} \right)\exp \left(- a \frac{d}{dx} \right) + G \left(\frac{x}{a} \right) \longrightarrow \phi(x) \frac{d}{dx} + \gamma(x).\nonumber \\
\label{e5}
\eea
Here we have imposed $F(x/a) + G(x/a) \rightarrow \gamma(x)$ and $aF(x/a) \rightarrow -\phi(x)$. This limit resembles the usual Darboux operator \cite{darboux1882, manas1997}. Moreover, our discrete $A$ in (\ref{e2}) is a particular case of a series of automorphisms, but we should stress that  $F(N) \neq \mbox{constant}$ is a more general choice and allows more freedom in our models. In connection with singularities in our limits, we should note that $a F(x/a)$ is regular at $a=0$, but $F(x/a)$ is not. The function $G(x/a)$ compensates for the singularity of $F(x/a)$ rendering a finite $\gamma(x)$. We shall come back to this point in connection with the P\"oschl-Teller potential as a plausible continuous limit.

With these considerations, we are ready to construct a discrete superpartner $\tilde H$ with the prescription

\bea
\tilde H = AA^{\dagger}= \tilde \Delta(N) T +  T^{\dagger} \tilde \Delta(N)^{\dagger} + \tilde V(N).
\label{e6}
\eea
The new potential and hopping functions are given by

\bea
\tilde \Delta(N) = G(N-1)^{\dagger} F(N), 
\label{e7}
\eea
\bea
\tilde V(N) =   G(N)G(N)^{\dagger} + F(N)F(N)^{\dagger},
\label{e8}
\eea
and their eigenvalues obey the relations $\tilde \Delta_n = G_{n-1}^* F_{n}, \tilde V_n = |G_{n}|^2 + | F_{n}|^2$. Remarkably, the functions which determine the superpartners come in pairs, since both $\Delta$ and $V$ must be modified.

\subsection{Isospectrality and Transparency \label{sec:3.1} }

Our method can be readily applied to bound states as well as scattering solutions. Let $|\psi_k\>$ be a solution of $H|\psi_k\>=E_k|\psi_k\>$. If $\tilde H|\tilde \psi_k\>= \tilde E_k| \tilde \psi_k\>$, we deduce the relations

\bea
|\tilde \psi_{k} \> = \left( E_{k+1} - E_0 \right)^{-1/2} \{ F(N)T + G(N) \} |\psi_{k+1} \>, \nonumber \\
\label{e9}
\eea
\bea
\tilde E_{k} = E_{k+1}.
\label{e10}
\eea
The $N=2$ supersymmetry is realized by defining supercharges with the help of $A, A^{\dagger}$ and the Pauli matrices $\sigma_{\pm}$. We have 

\bea
Q=\sigma_- A, \quad Q^{\dagger}=\sigma_+ A^{\dagger}, \quad \hcal = \{ Q, Q^{\dagger} \}. 
\label{e10.1}
\eea
where $\hcal$ is the central charge. The aforementioned isospectrality can be used for many purposes, but we are interested now in the transformation properties of the scattering matrix and their relation with transparency.

Let us define the conditions of a scattering problem in discrete variables. We take $H, \tilde H$ asymptotically periodic, such that in the limit $n \rightarrow \pm \infty$ we have $V_n \rightarrow V_{\pm}, \tilde V_n \rightarrow \tilde V_{\pm}, \Delta_n \rightarrow \Delta_{\pm}, \tilde \Delta_n \rightarrow \tilde \Delta_{\pm}$. The solutions with continuous parameter $k$ become Bloch waves: $\< n| \psi_k\> \rightarrow \ncal \mbox{e}^{ikn} $ and similarly for $\< n| \tilde \psi_k\>$. The limit values of $F$ and $G$ can be reconstructed via the relations $|F_{\pm}|^2= \ahalf (V_{\pm} \pm \sqrt{V_{\pm}^2-4\Delta_{\pm}^2})$, where $V_{\pm}^2 \geq 4 \Delta_{\pm}^2$. With this information on the factorization parameters, we are ready to apply the discrete Darboux transformation to the asymptotic form of the wave functions

\bea
\< n |A| \psi_k \> = \left( G_n + \mbox{e}^{ik} F_n \right) \psi_n^{k} \rightarrow  \left( G_{\pm} + \mbox{e}^{ik} F_{\pm} \right) \mbox{e}^{ikn}. \nonumber \\
\label{e11}
\eea
It may happen that the asymptotic regions at $\pm \infty$ have different associated constants, i.e. $V_+ \neq V_-, \Delta_+ \neq \Delta_-$. However, we always have the energy (or dispersion) relation $E = 2 \Delta_- \cos k + V_- = 2 \Delta_+ \cos k' + V_+$ satisfied at both ends of the array. In general $k'\neq k$, and the reflected and transmitted waves have different Bloch quasi-momenta, expressed by the limits

\bea
\<n | \psi_k \> \begin{array}{c} _{\longrightarrow} \\  _{n \rightarrow -\infty} \end{array} \mbox{e}^{ikn} + R \mbox{e}^{-ikn},
\label{e12}
\eea
\bea
\<n | \psi_k \> \begin{array}{c} _{\longrightarrow} \\  _{n \rightarrow +\infty} \end{array}  T \mbox{e}^{ik'n}.
\label{e13}
\eea
The application of the discrete Darboux transformation (\ref{e11}) to (\ref{e12}) and (\ref{e13}) leads to waves of the same energy but with modified transmission and reflection coefficients $\tilde T, \tilde R$. The results are

\bea
\tilde R = \left( \frac{G_- + \mbox{e}^{ik}F_-}{G_- + \mbox{e}^{-ik}F_-} \right) R
\label{e14}
\eea
\bea
\tilde T = \left( \frac{G_+ + \mbox{e}^{ik'}F_+}{G_- + \mbox{e}^{-ik}F_-} \right) T
\label{e15}
\eea
which constitute a generalization of the usual scattering matrix transformations with the replacement $ik \mapsto e^{ik}$. The reality condition for $G_{-}$ and  $F_-$ establishes that $|\tilde R|= |R|$. Moreover, in the problem of transparency $\tilde R$ vanishes if $R=0$.  Let us examine this possibility by proposing $H$ as a {\it free\ }hamiltonian, i.e. a periodic chain. This means that $\Delta_n=\Delta_{\pm} \equiv \Delta_0$ and $V_n=V_{\pm} \equiv V_0$. Now we must find $\tilde V_n$ and $\tilde \Delta_n$ by solving the recurrence 

\bea
|G_n|^2 = V_0 - |F_{n+1}|^2 = \frac{|\Delta_0|^2}{ |F_{n}|^2}
\eea 
for $|F_n|^2$. With the definitions $\mu_{\pm} = \ahalf (V_{0} \pm \sqrt{V_{0}^2-4\Delta_{0}^2})$ we obtain

\bea
F_n = \sqrt{\frac{\alpha_+ \mu_+ ^{n} + \alpha_- \mu_- ^{n}}{\alpha_+ \mu_+ ^{n-1} + \alpha_- \mu_- ^{n-1}}},  
\label{e16}
\eea
\bea
G_n = \Delta_0 \sqrt{\frac{\alpha_+ \mu_+ ^{n-1} + \alpha_- \mu_- ^{n-1}}{\alpha_+ \mu_+ ^{n} + \alpha_- \mu_- ^{n}}}, 
\label{e17}
\eea
where $\alpha_{\pm}$ are arbitrary constants of the same sign, preserving the reality of $F_n$. The potential and hopping functions are now

\bea
\tilde V_n = V_0  + \frac{\alpha_+ \mu_+ ^{n} + \alpha_- \mu_- ^{n}}{\alpha_+ \mu_+ ^{n-1} + \alpha_- \mu_- ^{n-1}} - \frac{\alpha_+ \mu_+ ^{n+1} + \alpha_- \mu_- ^{n+1}}{\alpha_+ \mu_+ ^{n} + \alpha_- \mu_- ^{n}} \nonumber \\ 
\label{e18}
\eea
\bea
\tilde \Delta_n = \Delta_0 \frac{\sqrt{ (\alpha_+ \mu_+ ^{n} + \alpha_- \mu_- ^{n})(\alpha_+ \mu_+ ^{n-2} + \alpha_- \mu_- ^{n-2}) } }{ \alpha_+ \mu_+ ^{n-1} + \alpha_- \mu_- ^{n-1} }.\nonumber \\ 
\label{e19}
\eea
Finally, $\tilde H$ given in (\ref{e6}) is a non-trivial monoparametric familiy enjoying the property of being reflectionless. One can show that the solutions depend only on $\gamma \equiv \alpha_- / \alpha_+$, and that $\tilde V, \tilde \Delta$ can be put in terms of hyperbolic functions $\sinh ( n \ln \mu_{\pm}), \cosh ( n \ln \mu_{\pm})$. In figure \ref{fig:1} we show the behavior of potentials and hopping parameters under the modification of $\gamma$, resulting in a translation of the interaction region or potential well. This is reminiscent of shape invariance in continuous variables, where generalized P\"oschl-Teller or Scarf potentials can be translated at will on the real line (among other operations, such as rescaling). However, such a similarity is to be taken with a grain of salt, since continuous translations of discrete variables yield the same (translated) potentials only if $\nu = \ln(\gamma) / \ln (\mu_+ / \mu_-)$ is an integer. We can distinguish these features in the panels of figure \ref{fig:1}, as the centers of $\tilde V, \tilde \Delta$ move to the right and the functions suffer slight variations for non-integer values of $\nu$.

\begin{figure*}[t]
\includegraphics[width=10.6cm]{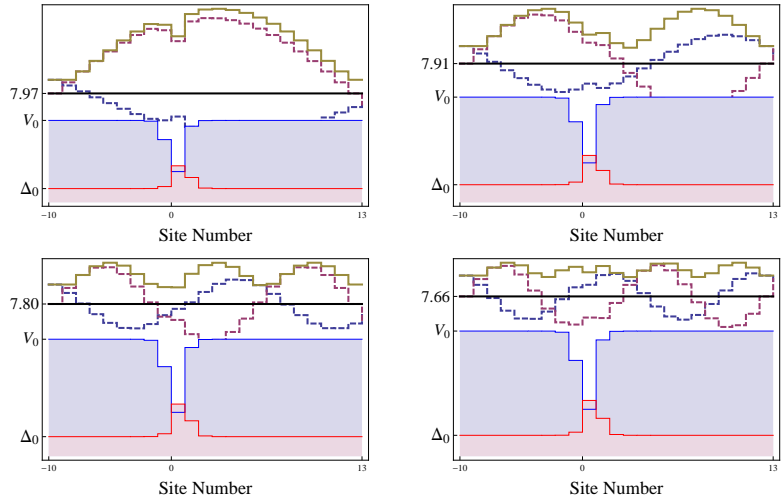}
\caption{\label{fig:2} Numerical solutions of the scattering problem in discrete variable $n$ for a Bloch wave impinging from the left. Each panel shows a different scattering energy $E$. The parameters are $\gamma=\mbox{e}^{-6}$, $V_0 = 6 \Delta_0$. Solid gold: $|\tilde \psi|$, dashed blue:  $\mbox{Re}(\tilde \psi)$,  dashed red:  $\mbox{Im}(\tilde \psi)$, blue-filled curve: $\tilde V$, red-filled curve: $\tilde \Delta$, solid black: $E$. In all cases $|T|^2 = 1$ and the original Bloch wave picks up a phase at the right end of the array.}
\end{figure*}

\subsection{The continuous limit \label{sec:3.2}}

It is important to make contact with the well-known results of transparent potentials in continuous SUSYQM. The limits can be reached by letting $a \rightarrow 0$ as before. In order to recover a Schr\"odinger equation with double derivatives and ground state energy $U_0$, we must impose $x=an$, $\Delta_0 \sim - R_0 / a^2$ with $R_0 >0$ and $V_0 + 2\Delta_0 \sim U_0$. In the process, we note that $T \sim 1 - a\cdot d/dx$ and $(\mu_+ / \mu_-)^n \sim \exp \left(4x\sqrt{R_0/U_0} \right)$. The particular choice $\gamma=1$ leads to a familiar case of hyperbolic superpotentials; we have

\bea
F_n \sim \frac{\sqrt{R_0}}{a} + \sqrt{\frac{U_0}{R_0}} \tanh \left( 2\sqrt{\frac{U_0}{R_0}} x \right),
\label{e21}
\eea
\bea
G_n \sim -\frac{\sqrt{R_0}}{a} + \sqrt{\frac{U_0}{R_0}} \tanh \left( 2\sqrt{\frac{U_0}{R_0}} x \right),
\label{e22}
\eea
and the Darboux operator becomes

\bea
A \sim 2\sqrt{\frac{U_0}{R_0}} \tanh \left( 2\sqrt{\frac{U_0}{R_0}} x \right) + \sqrt{R_0} \frac{d}{dx}.
\label{e23}
\eea
This is the usual operator for the Rosen-Morse superpotential with non-zero ground state $U_0$.

\subsection{A numerical test for transparency \label{sec:4}}
We test the reflectionless property by solving numerically the scattering problem for various energies. A reasonable choice of parameters for the potential and hopping functions is $\gamma=e^{-6}$ (producing strong asymmetry in the potential) and $V_0/\Delta_0 = 6$, localizing the region of interaction in a small portion of a few sites. The numerical solution is reached by imposing a Bloch wave at least at two sites, since (\ref{a4}) is a second order recurrence. For instance, at sites $n=0$ and $n=1$ we have

\bea
\tilde \psi_0 = \mbox{e}^{i \times 0}= 1,\quad \tilde \psi_1 = \mbox{e}^{ik}.
\eea
Such boundary conditions generate $\tilde \psi_{n+1}$ through the recurrence
\bea
\tilde \psi_{n+1} = \frac{ \tilde \Delta_n \tilde \psi_{n-1} +  (E-\tilde V_n)\tilde \psi_n}{\tilde \Delta_{n+1}},
\label{e20}
\eea
provided that $E$ and $k$ are related by $E=V_0 + \Delta_0 \cos k$. We have used four different energies in the scattering regime: $E/\Delta_0 = 7.66, 7.80, 7.91, 7.97$, verifying that the modulus of the {\it transmitted\ }wavefunction recovers the value $1$ in all cases. The results are shown in figure \ref{fig:2}, where the modulus, the real and the imaginary part of $\psi_n$ are displayed. At the right end of the array (25 sites) the wave recovers its modulus and phase factor, but at this region the potential is negligible and the solution will continue to be a Bloch wave propagating to the right if the array is prolonged indefinitely.

\begin{figure*}[t]
\includegraphics[width=10.6cm]{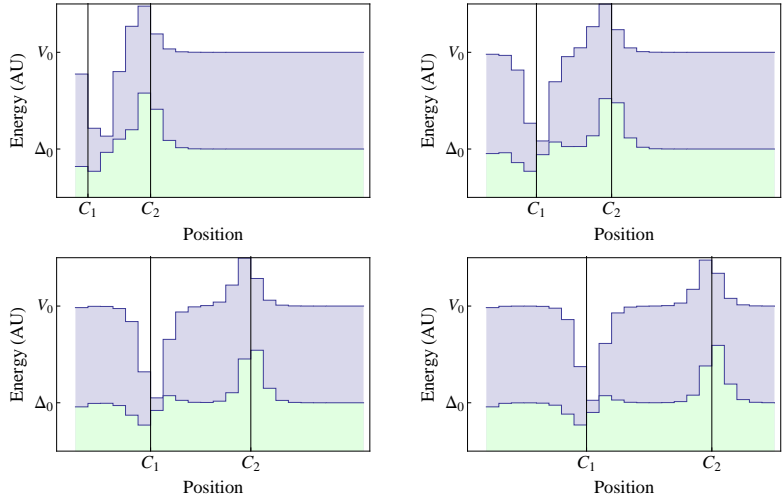}
\caption{\label{fig:3} The motion of discrete solitonic potentials (blue-filled curve) and solitonic hopping functions (green-filled curve). The parameter $\alpha = 1.5$ produces two solitons (a maximum and a minimum) for each graph. As $\gamma$ increases, their centers $C_1$ and $C_2$ move to the right at different velocities. This is shown progressively from top left to right bottom.}
\end{figure*}

\subsection{Biparametric solitons \label{sec:5}}

The solutions of the Korteweg-deVries (KdV) equation \cite{kruskal1967, kdv1895} are known to be represented by a hierarchy of superpotentials \cite{wang1990}. Such a hierarchy can be obtained through a step-by-step method for generating superpartners. In our discrete case, however, the application of new Darboux transformations can be increasingly challenging. Yet, a simpler strategy to obtain families of solutions consists of finding at once all the superpartners of a given reflectionless problem. We proceed in this direction in what follows. Let us start with a monoparametric transparent problem given by (\ref{e18}) and (\ref{e19}). We now consider a hamiltonian

\bea
H^{(\alpha)} = \Delta^{(\alpha)}(N) T + T^{\dagger} \Delta^{(\alpha)}(N)^{\dagger} + V^{(\alpha)}(N),
\label{e24}
\eea
where $\alpha$ is a new parameter, yet to be determined. The factorization procedure yields the relations

\bea
V^{(\alpha)}_n = |G_n^{(\alpha)}|^2 + |F_{n+1}^{(\alpha)}|^2, \nonumber \\ \Delta^{(\alpha)}_n = \left[G_n^{(\alpha)}\right]^* F_n^{(\alpha)},
\label{e25}
\eea
but another set of recurrences in terms of $\tilde V_n$ and $\tilde \Delta_n$ must be satisfied for the reflectionless problem:
\bea
\tilde V_n = |G_n^{(\alpha)}|^2 + |F_{n}^{(\alpha)}|^2, \nonumber \\ \tilde \Delta_n = \left[G_{n-1}^{(\alpha)}\right]^* F_n^{(\alpha)},
\label{e25.1}
\eea
where (\ref{e18}) and (\ref{e19}) must be substituted in the l.h.s. of (\ref{e25.1}). These relations are sufficient to determine $|F_n^{(\alpha)}|^2$ and $|G_n^{(\alpha)}|^2$. A particular solution is given, of course, by (\ref{e16}) and  (\ref{e17}), but the most general solution of (\ref{e25}) is a continued fraction

\bea
|F_n^{(\alpha)}| = \left[ |\tilde \Delta_n|^2,  \tilde V_{n-1}; ...; |\tilde \Delta_1|^2, \tilde V_0 - \alpha \right], \nonumber \\
|G_n^{(\alpha)}| = \frac{|\tilde \Delta_{n+1}|^2}{\left[ |\tilde \Delta_{n+1}|^2,  \tilde V_{n}; ...; |\tilde \Delta_1|^2, \tilde V_0 - \alpha \right]}.
\label{e26}
\eea
We identify the new parameter $\alpha$ with the initial condition of the recurrence, i.e. $\alpha = |F_0^{(\alpha)}|^2$. The potentials and hopping functions can be reconstructed by means of the relations

\bea
V_n^{(\alpha)} = \tilde V_n + |F_{n}^{(\alpha)}|^2 - |F_{n+1}^{(\alpha)}|^2 ,
\label{e27}
\eea
\bea
\Delta_n^{(\alpha)} = \tilde \Delta_n \frac{F_{n}^{(\alpha)}}{F_{n+1}^{(\alpha)}}.
\label{e28}
\eea
The expressions (\ref{e27}) and (\ref{e28}) represent a biparametric family of transparent potentials and hopping functions. It is worthwhile to investigate their behavior as a function of both $\alpha$ and the original parameter $\gamma$. For example, setting $\alpha=1.5$ takes us to two solitons for each of the functions $V^{(\alpha)}_n, \Delta^{(\alpha)}_n$. For values $\gamma \sim 1$ the solitons are close to each other (see figure \ref{fig:3}). Increasing $\gamma$ exponentially produces their motion with respect to the origin as well as a relative displacement between them. Thus, we have two solitons with two different velocities \cite{wang1990}.

\section{Conclusion and outlook \label{sec:6}}

Lattice design by site and coupling engineering gives rise to many possibilities of which the present paper is an example. A concrete experiment showing transparency --among other properties predicted by SUSYQM-- can be proposed using current technologies, as discussed in section \ref{sec:experiment}. A particular configuration using dielectric slabs was provided in \ref{sec:slabs}. We also recognize that SUSYQM is indeed a powerful method; its application to discrete problems has been demonstrated by finding systems with a desired property. From the mathematical point of view, we have found that some aspects of solitons \cite{lederer2008, nijhoff1995, wang1990} can be reproduced also in tight-binding arrays, motivating further explorations towards discrete spinorial KdV equations. The extension of the present study to 2D lattices seems plausible. Moreover, discrete exactly solvable problems \cite{natig1991, natig2001} and their relation with shape invariant potentials can be explored in this context.

\begin{acknowledgments}
I am grateful to T. H. Seligman for useful comments on the manuscript. Financial support from CONACyT under project CB 2012-180585 is acknowledged. 
\end{acknowledgments}



\providecommand{\noopsort}[1]{}\providecommand{\singleletter}[1]{#1}%

\end{document}